\documentclass[a4paper]{jpconf}
\usepackage{graphicx}

\newcommand{\Jpsi} {\mbox{J\kern-0.05em /\kern-0.05em$\psi$}}
\newcommand{\nY}{$\Upsilon$\ }
\newcommand{\npi}{$\pi$\ }
\newcommand{\nL}{$\Lambda$\ }
\newcommand{\npt}{$\rm p_T$ }
\newcommand{\raa}{$\rm R_{AA}$ }
\begin{document}
\title{Results from the first heavy ion run at the LHC}

\author{J. Schukraft\footnote{
Invited talk  at the Rutherford Centennial Conference
on Nuclear Physics, July 25 - 29, 2011, Manchester, UK.
}}

\address{CERN, Div. PH, 1211 Geneva 23}

\ead{schukraft@cern.ch}

\begin{abstract}
Early November 2010, the LHC collided for the first time heavy ions, Pb on Pb, at a centre-of-mass energy of 2.76 TeV/nucleon. This date marked both the end of almost 20 years of preparing for nuclear collisions at the LHC, as well as the start of a new era in ultra-relativistic heavy ion physics at energies exceeding previous machines by more than an order of magnitude. This contribution summarizes some of the early results from all three experiments participating in the LHC heavy ion program (ALICE, ATLAS, and CMS), which show that the high density matter created at the LHC, while much hotter and larger, still behaves like the very strongly interacting, almost perfect liquid discovered at RHIC. Some surprising and even puzzling results are seen in particle ratios, jet-quenching, and Quarkonia suppression observables. The overall experimental conditions at the LHC, together with its set of powerful and state-of-the-art detectors, should allow for precision measurements of quark-gluon-plasma parameters like viscosity and opacity.
\end{abstract}

\section{Introduction}
The aim of ultra-relativistic heavy ion physics (URHI) is the study of matter under extreme conditions of temperature and density, where QCD predicts the formation of a new phase of matter, the quark-gluon plasma (QGP). In the QGP, partons are deconfined -- i.e. not bound into composite colourless hadrons -- and chiral symmetry is (approximately) restored -- i.e. the light quarks are (approximately) mass less. The mission of URHI is then to \emph{search} for the QGP, to \emph{measure} its properties, and along the way to \emph{discover} QCD in the non-pertubative sector, where the strong interaction is strong indeed.

After ten years at RHIC (Relativistic Heavy Ion Collider) at $\sqrt{s}$ up to 200 GeV/nucleon (and a comparable time at fixed target machines at one tenth this energy), the \emph{search} for the QGP is essentially over and its \emph{discovery} is well under way~\cite{Heinz:2000bk, rhichwhitepaper}.
A 'Heavy Ion Standard Model' (HISM) has emerged which characterises the high density state created in nuclear collisions as an extremely strongly interacting and almost perfect fluid, sometimes called the 'sQGP' (where the 's' stands for 'strongly interacting'). Describing it as a 'fluid' implies 'macroscopic matter' properties and collective degrees of freedom, with dimensions significantly larger than the mean free path and existing for a time significantly longer than the relevant relaxation times. This fluid is almost opaque and absorbs much of the energy of any fast parton (quark or gluon) which travels through -- a process referred to as 'jet quenching' -- and it reacts to pressure gradients by flowing almost unimpeded and with very little internal friction (i.e. very small viscosity).  However, the precision \emph{measurement} of QGP parameters has just begun.

\subsection{The role of LHC after RHIC}
The heavy ion program at LHC may be divided roughly into three different phases: {\bf 1) Quantifying differences:} At the high centre-of-mass energy of LHC, the matter created should be significantly different in terms of energy density, lifetime, and freeze-out volume compared to RHIC. In addition, jets and heavy quarks will be copiously produced making these (and other) 'hard probes' an important tool to measure QGP properties.
{\bf 2) Test and validate the 'Heavy Ion Standard Model':} The HISM has emerged less than 10 years ago from RHIC and can now for the first time be probed in a different energy regime to see if it is robust enough to make reliable extrapolations and predictions or if we find surprises which require additions or changes to the HISM. {\bf 3) Precision measurements:} Finally, a quantitative and systematic study of the sQGP should be carried out at the LHC to measure, with some precision, its properties and parameters like viscosity, opacity, Debye screening mass, and the equation-of-state.

The LHC heavy ion program will need a decade or more to go through this program, but already the first year results have made significant progress on points 1) and 2) above. Precision measurements are still some way ahead, but first indications from collective flow, jet-quenching and quarkonia suppression measurements are very promising in this respect.

\section{Quantifying differences with respect to lower energies}

\begin{figure}[!t]
\centerline{\includegraphics[width=1.0\textwidth]{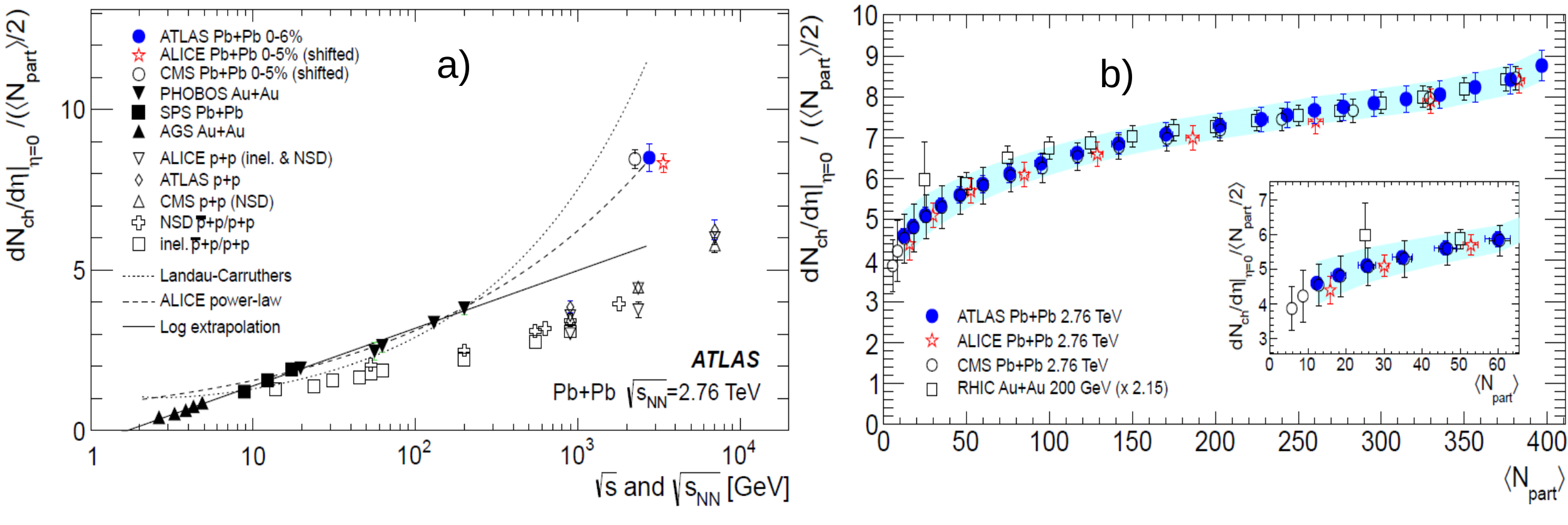}}
\caption{a): Charged particle pseudorapidity density ${\rm d}N_{\rm ch}/{\rm d}\eta$ per colliding nucleon pair ($0.5 N_{\rm part}$) versus centre of mass energy for pp and AA collisions. b): ${\rm d}N_{\rm ch}/{\rm d}\eta$ per colliding nucleon pair versus centrality as measured by the number of nucleons participating in the collision ($N_{\rm part}$). Figures from ref~\cite{ATLASmult}. }
\label{figpb1}
\end{figure}

The first, and eagerly awaited, result from LHC concerned the charged particle multiplicity density in central collisions. The value finally measured with Pb--Pb at 2.76 TeV/nucleon~\cite{Aamodt:2010pb}, ${\rm d}N_{\rm ch}/{\rm d}\eta \approx 1600$, was somewhat on the high side of more recent (post-RHIC) predictions~\cite{lhcpred} but well below the maximum values ($>4000$) anticipated during the design phase of LHC in the early 90's. From the measured multiplicity in central collisions one can derive a rough estimate of the energy density, which turns out to be at least a factor three above RHIC. The corresponding increase of the initial temperature is about 30\%, even with the conservative assumption that the formation time (when a thermal equilibrium is first established) does not decrease from RHIC to LHC. This is indeed matter under extreme conditions, reaching a temperature which is 200,000 times hotter than the centre of our sun and an energy density fifty times larger than in the core of a neutron star!

When combined with lower energy data, the charged particle production per participant pair, $(dN_{ch}/d\eta)/(0.5 N_{part})$, is seen in 
Fig.~\ref{figpb1}a to rise with centre of mass energy $\sqrt s$ approximately like $s^{0.15}$, stronger than in pp ($s^{0.11}$). Somewhat surprising is the fact that the centrality dependence of particle production (Fig.~\ref{figpb1}b)~\cite{Collaboration:2010cz, ATLASmult}
is practically identical to the one of Au-Au at RHIC, at least for $N_{\rm part} >50$, given that the impact parameter dependent nuclear modifications is expected to be much stronger at LHC. Indeed, models with either strong nuclear modifications or different saturation-type calculations ('Colour Glass Condensate' based models) do describe the impact parameter dependence best.

The agreement between the three LHC experiments shown in Fig.~\ref{figpb1} is excellent, the differences being on the level of a few percent at most and smaller than the quoted systematic errors. This is all the more remarkable given that the first multiplicity measurements emerged within days of the first collisions, and that a similar agreement between RHIC experiments took several years to achieve.

\begin{figure}[!t]
\centerline{\includegraphics[width=1.0\textwidth]{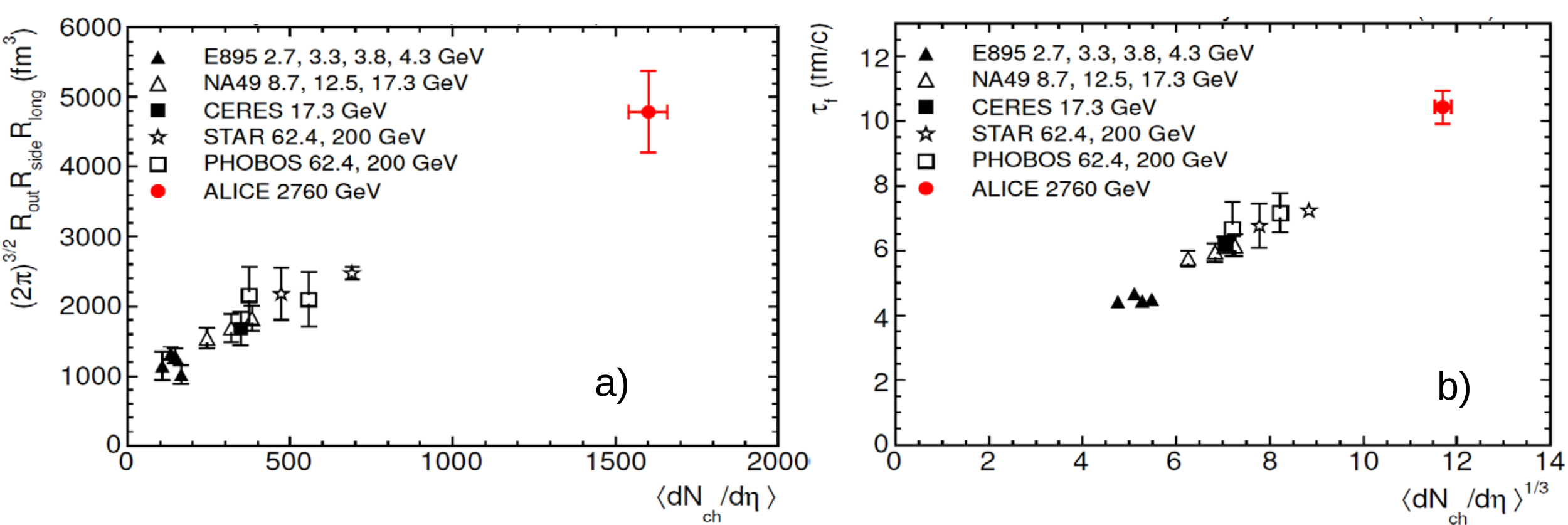}}
\caption{ a): Local freeze-out volume as measured by identical pion interferometry at LHC compared to central gold and lead collisions at lower energies. b): The system lifetime (decoupling time) $\tau_{f}$ compared to results from lower energies. 
Figures from ref~\cite{Aamodt:2011mr}.
}
\label{figpb2}
\end{figure}

The freeze-out volume (the size of the matter at the time when strong interactions cease) and the total lifetime of the created system (the time between collision and freeze-out) was measured with identical particle interferometry (also called Hanbury-Brown Twiss or HBT correlations)~\cite{Aamodt:2011mr}. Compared to top RHIC energy, the local 'comoving' freeze-out volume (Fig.~\ref{figpb2}a) increases by a factor two (to about 5000 fm$^3$) and the system lifetime (Fig.~\ref{figpb2}b) increases by about 30\% (to 10 fm/c), pretty much in line with the predictions of the HISM~\cite{lhcpred}.  

\section{Testing the Heavy Ion Standard Model}
\subsection{Elliptic and radial flow}
The most critical test of the HISM comes from the measurement of the elliptic flow at LHC, the pillar which supports the 'fluid' interpretation of the QGP. Flow refers to a correlation between space and momentum variables, i.e. particles close in space show similar velocity in both magnitude and direction. Such a collective behaviour is in contrast to random thermal motion where space and momentum variables are in general not correlated. Different flow patterns are observed in heavy ion collisions and quantified in terms of a Fourier decomposition. The second harmonic is called 'elliptic' flow and the zero order (i.e. isotropic) harmonic is called 'radial' flow. Other harmonic components (e.g. 1st, 3rd,..), can be measured as well.

\begin{figure}[!t]
\centerline{\includegraphics[width=1.0\textwidth]{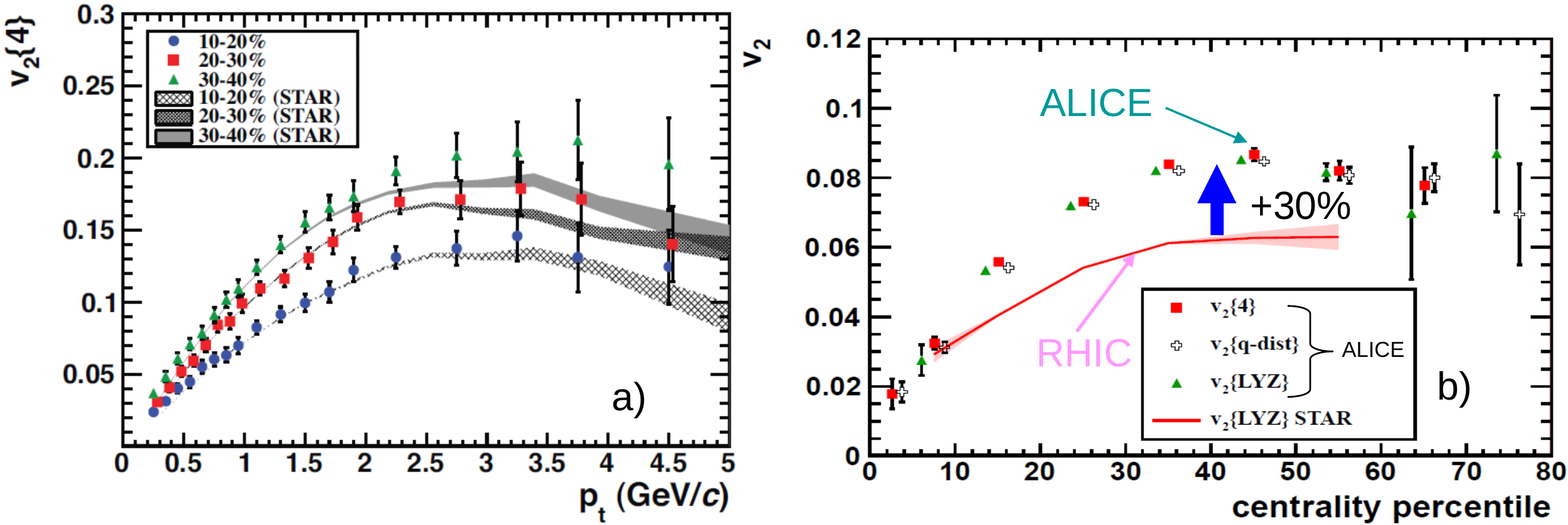}}
\caption{ a): Elliptic flow coefficient $v_2$ as a function of transverse momentum for different centrality ranges. Symbols ALICE/LHC, shaded bands STAR/RHIC. b): Integrated elliptic flow, measured with several methods, as a function of collision centrality.
Figures adapted from ref~\cite{Aamodt:2010pa}.
}
\label{figpb3}
\end{figure}

Collective observables like particle flow are usually described in the framework of hydrodynamic models, where initial conditions (e.g. geometry and pressure gradients) and fluid properties (e.g. viscosity and equation-of-state) fully determine the observed pattern of collective motions. Assuming no or only small changes in the fluid properties between RHIC and LHC, hydrodynamic models predict firmly that the elliptic flow coefficient $v_2$, measured as a function of $p_T$, should change very little, whereas the momentum integrated flow values should either stay approximately constant or rise by at most some 30\%. As shown in Fig.~\ref{figpb3}, this predictions were confirmed very quickly~\cite{Aamodt:2010pa}. The $p_T$ differential flow (Fig.~\ref{figpb3}a) is unchanged whereas the $p_T$ integrated flow (Fig.~\ref{figpb3}b) increases by some 30\%, at the upper edge but still within predictions from hydro. The increased integrated flow is linked to the average transverse momentum, which is significantly higher at LHC than at RHIC. While the average $p_T$ increases also in pp with energy, because hard  and semi-hard processes become more important, hydrodynamics predicts in nuclear collisions an increase in the radial flow velocity leading to a characteristic $p_T$ and mass dependence of the spectra.

The characteristic mass dependent 'blue shift' of radial flow can be seen in Fig.~\ref{figpb4}a, which shows the $p_T$ spectra of identified pions, kaons and anti-protons~\cite{jsqm}.
The spectral slopes change significantly compared to RHIC, most dramatically for protons, indicative of a much stronger radial flow. A first attempt to extract radial flow with a hydro inspired fit function indicates that the average flow velocity reaches about 2/3 of the speed of light (meaning that the leading edge expands collectively at essentially the speed of light). Also the radial flow is found to be at the upper edge, but within, expectations, and is therefore fully consistent with the elliptic flow magnitude mentioned earlier.

\subsection{Particle ratios}
Particle production (\npi, K, p, \nL,..) is a non-pertubative process and can in general not be predicted by QCD. In phenomenological, QCD inspired Monte Carlo event generators, particle ratios have to be fitted with a large number of 'ad hoc' parameters. Surprisingly however, and without 'a priori' justification, particle ratios can be described with good precision, of order 10-20\%, in a very simple thermal/statistical picture, which assumes that particles are created in thermal (or phase space) equilibrium governed by a scale parameter T, referred to as 'temperature'. Production of a particle with mass $m$ is then essentially suppressed by a Boltzmann factor $e^{-m/T}$. Conservation laws introduce additional (but not arbitrary) parameters, like the baryochemical potential $\mu_B$ which accounts for baryon number conservation. As particles containing strange quarks are produced less abundant in pp collisions then predicted in these models, an additional 'ad hoc' parameter $\gamma_s$ has to be introduced to capture this strangeness suppression. The temperature is found in all high energy collisions (pp, $e^+e^-$, AA) to be close to 160 MeV, while $\gamma_s$ increases from 0.5-0.6 in pp to 0.9-1 in AA. The disappearance of strangeness suppression (usually called strangeness enhancement) in nuclear collisions was one of the first signals predicted for the QGP, and today the fact that all stable or long lived particles are produced in heavy ion reactions with 'thermal ratios' is considered to be an established fact and indeed a pillar of the HISM.

\begin{figure}[!t]
\centerline{\includegraphics[width=1.0\textwidth]{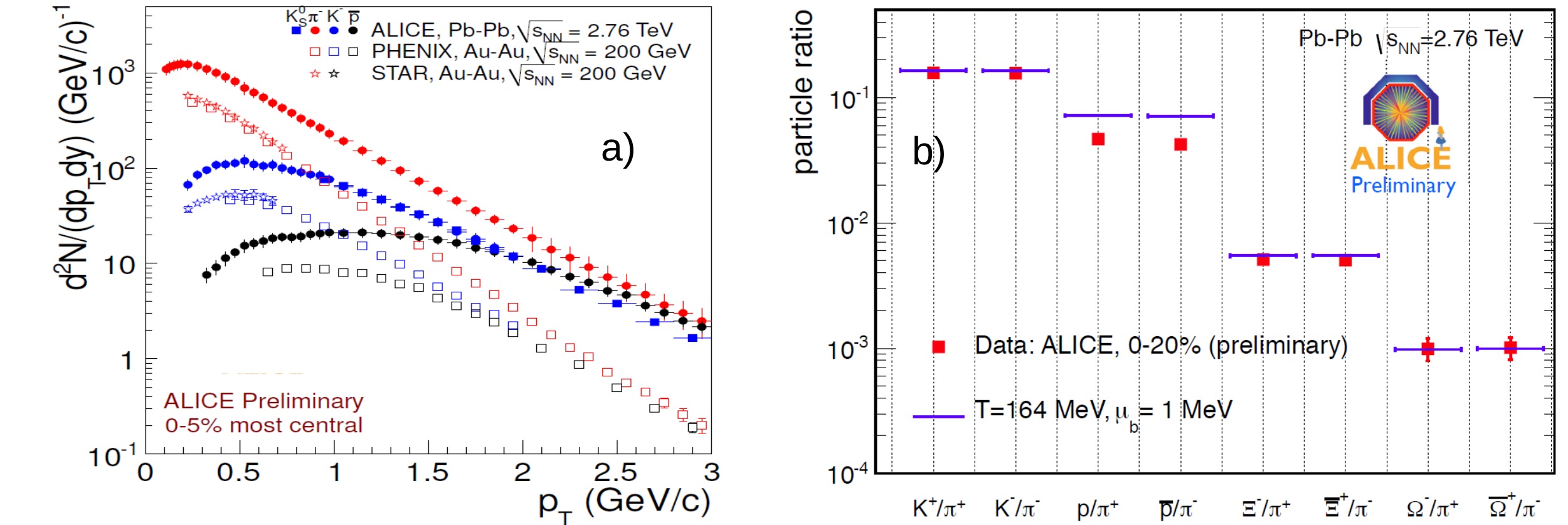}}
\caption{ a): Transverse momentum spectra of identified particles from LHC (closed symbols) and RHIC (open symbols). b): Particle ratios measured in central (0-20\%) collisions. The full line is the prediction from a statistical model~\cite{Andronic}.
}
\label{figpb4}
\end{figure}

It therefore came as a complete surprise when first particle ratios at the LHC~\cite{jsqm} showed that protons are too low compared to the prediction by a factor of 1.5 at least. Fig.~\ref{figpb4}b shows the ratio relative to pions for a number of particle species, including multi-strange baryons. The prediction of a thermal model~\cite{Andronic} is shown as a line, using the canonical conditions expected at the LHC (T $\approx 165$ MeV, $\mu_B \approx 0,  \gamma_S = 1$). While strange and multi-strange particles are very well described, indicating that strangeness is not suppressed also at LHC (i.e. $\gamma_S = 1$), protons are strikingly off, well outside the usual precision of the thermal model. Before concluding that something is wrong or missing in current implementations of the thermal model, the data, which are still preliminary, and various corrections (e.g. feed-down from weak decays) have to be thoroughly checked. Also the proton ratios at RHIC should be revisited, as there are indications of a smaller, but still significant, tension between model fits and data.

\subsection{$\Jpsi$ and \nY suppression}

\begin{figure}[!t]
\centerline{\includegraphics[width=1.0\textwidth]{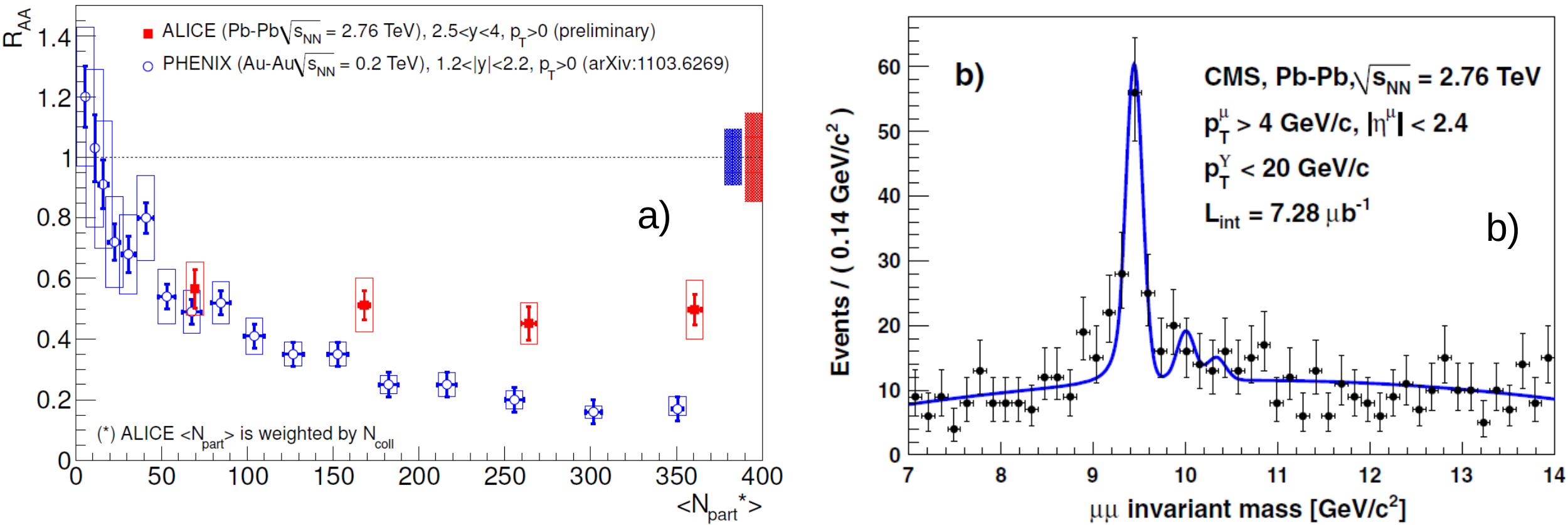}}
\caption{ a): Nuclear modification factor \raa versus centrality ($N_{part}$). Closed symbols from ALICE/LHC, open symbols from PHENIX/RHIC. b): Dimuon invariant-mass distributions in the \nY region~\cite{Chatrchyan:2011pe}; the solid line is a fit to extract the yields. 
}
\label{figpb5}
\end{figure}

$\Jpsi$ production, the classical deconfinement signal, has confounded expectations and interpretations ever since the first nuclear suppression, measured with Oxygen beams at the SPS, was announced in 1987. The puzzling fact that $\Jpsi$ suppression is found to be rather similar in magnitude at SPS and RHIC may indicate that either only $\psi$' and $\chi_c$ (which both decay into $\Jpsi$) are suppressed at both SPS and RHIC, but not the $\Jpsi$ itself, or that direct $\Jpsi$ suppression is indeed present at RHIC but more or less balanced by enhanced secondary $\Jpsi$ production via coalescence of two independently created charm quarks. A resolution to this puzzle should come from measuring the $\Jpsi$ at LHC (where coalescence effects will be stronger because of the more abundant charm production, eventually even leading to $\Jpsi$ enhancement) and by comparing the suppression patterns of the $\Jpsi$ and $\Upsilon$ families.

The nuclear modification factor \raa for the $\Jpsi$, which quantifies the measured over the expected production ratio (\raa = rate measured with nuclei divided by the properly normalised rate measured in pp) has been measured by ALICE down to zero \npt as a function of centrality (Fig.~\ref{figpb5}a). The \raa value of about 0.5, with very little centrality dependence, is almost a factor 2 larger than the one measured by PHENIX at comparable forward rapidity; the difference is less but still significant when comparing with PHENIX results at midrapidity. The suppression is also significantly less than the one reported by ATLAS and CMS~\cite{:2010px, Silvestre:2011ei}. However, the ATLAS/CMS results are for high \npt and closer to midrapidity, which would indicate that the suppression either increases with transverse momentum or decreases with rapidity.

While these results are very intriguing, and would indicate regeneration via coalescence if taken at face value, there are still a number of unknowns to be clarified before any firm conclusions can be drawn. In particular the influence of shadowing/saturation effects on nuclear structure functions must be taken into account, as well as 'cold matter' absorption which may (or may not) be important at the LHC energy. For this reason a p--Pb run, which will address these nuclear effects, will be needed and is under discussion.

A first measurement of \nY production with Pb--Pb at LHC~\cite{Chatrchyan:2011pe} is shown in Fig.~\ref{figpb5}b. While statistical errors are still large, there is a clear indication that the excited, higher mass \nY states are more suppressed then the tightly bound ground state $\Upsilon(1S)$, as expected in a deconfinement scenario. Shadowing effects are probably less important for the \nY than for the $\Jpsi$ but should still be quantified via pA reactions. With the order of magnitude increase in luminosity expected for the $2^{nd}$ ion run in 2011, and additional input from proton-nucleus running in 2012, a complete picture for the charm and bottom -onia states should become available and hopefully help lift the fog which still obscures a clear interpretation of the observed suppression patterns from SPS to LHC.

\section{Towards precision measurements}

\begin{figure}[!t]
\centerline{\includegraphics[width=1.0\textwidth]{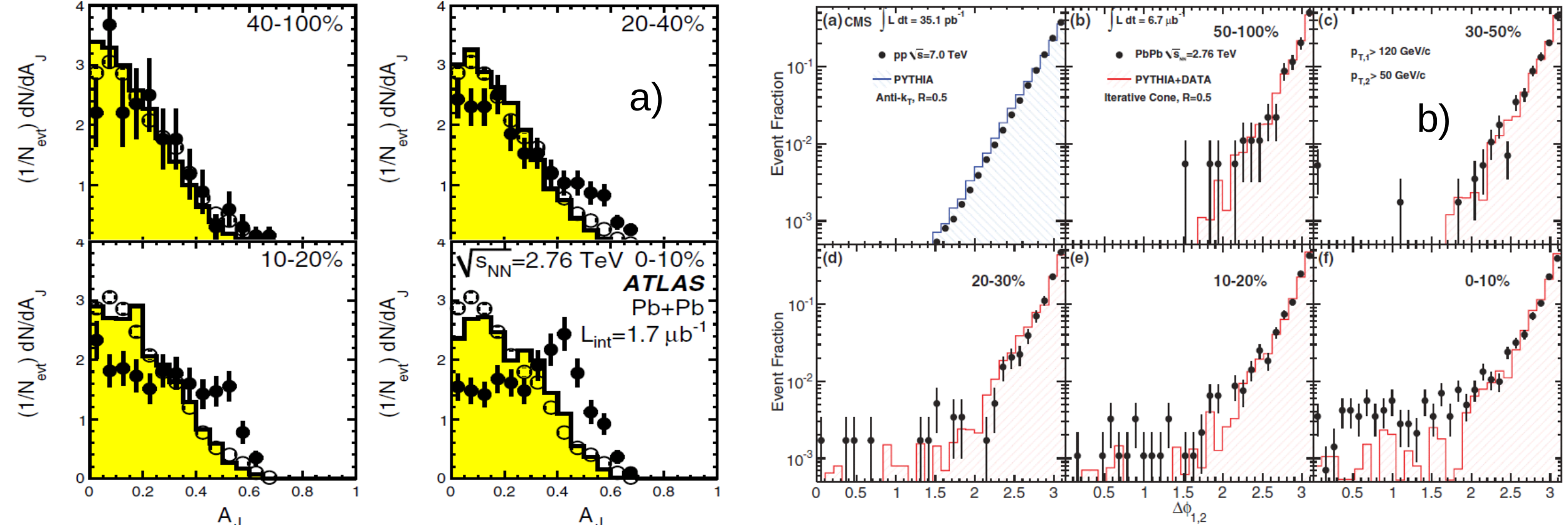}}
\caption{ a): Dijet asymmetry $A_j = (p_{T1} - p_{T2})/(p_{T1} + p_{T2})$  for different centrality selections for data (points) and Monte Carlo (histogram)~\cite{Aad:2010bu}. Proton-proton
data (7 TeV) are shown as open circles. b): Azimuthal angle $\Delta \phi_{1,2}$ between dijets for different centrality selections~\cite{Chatrchyan:2011sx}. Data are shown as black points,
while the histograms show Monte Carlo (PHYTHIA) events without jet quenching.
}
\label{figpb7}
\end{figure}

With the heavy ion standard model having passed its first tests fairly well (particle ratios however need to be sorted out), the programme of precision measurements is now starting at the LHC. 
A major advance in quantifying the shear viscosity, which at RHIC was measured to be within a factor 3-4 of the conjectured quantum limit for a perfect fluid, will require a good estimate of remaining non-flow contributions, as well as a better constraint on the initial conditions, i.e. the geometry of the collision zone (and its fluctuations) which drive the various flow components. Progress is already being made on both fronts, in particular by looking in more detail at the centrality dependence and at higher Fourier components ($v_3, v_4,..$)~\cite{:2011vk,Qiu:2011hf}, as will be discussed in detail by R. Snellings elsewhere in these proceedings.

\subsection{Jet quenching}
The energy advantage of the LHC is most evident in the area of parton energy loss (or jet quenching, the 'opaque' aspect of the sQGP), where the kinematic reach vastly exceeds the one available at RHIC. With high $p_T$ jets easily visible above the soft background, jet quenching is qualitatively evident at LHC already by visual inspection of jets where a striking energy imbalance develops for central collisions~\cite{Aad:2010bu}. 

In a quantitative analysis one has to extract how much energy is actually lost per unit path length (i.e. measure the transport coefficient), how and where it is lost (multiple soft versus few hard scatterings), and how this energy loss depends on a number of variables. Depending on the energy loss mechanism, it may depend linearly, quadratic, or even to the $3^{rd}$ power on the path length; because of the respective colour charges it should be stronger for gluons than for quarks; it should even depend on the quark mass with light quarks ($p_T/m >> 1$) losing more energy than heavy ones.  Eventually, an answer to all these questions will take several years (and pA comparison data as well), but significant steps have already been taken with the first year data.

ALICE has measured the nuclear modification of charged particle momentum distributions out to 20 GeV~\cite{Aamodt:2010jd}, where the spectra are dominated by leading jet fragments. The \raa ratio reaches a minimum at around 6 GeV, but then rises again smoothly towards higher momentum; a feature which was qualitatively predicted by some models for LHC~\cite{lhcpred}. Unlike at RHIC, where \raa is compatible with being flat above 5-6 GeV, the extended kinematic reach and the high \npt rise permit a better discrimination between different jet-quenching models and give a better constraint for the transport coefficient (energy loss parameter)~\cite{Renk:2011vn}.

ATLAS and CMS have measured fully reconstructed jets above 50-100 GeV jet energy. Fig.~\ref{figpb7}a shows the dijet asymmetry ($p_T$ difference divided by \npt sum) for peripheral (top right) up to central (bottom left) collisions. For peripheral collisions, the nuclear distribution (closed symbols) peaks close to zero, i.e. balanced dijets, and is similar to the one in pp (open symbols) and to the one expected from event generators without jet quenching (histogram). For central collisions, a marked asymmetry develops with unbalanced jets ($A_j > 0$) now more frequent than balanced ones. By comparing this distribution with a jet quenching calculation, the average energy loss has been estimated to be of the order of 10-20 GeV (with a very wide distribution, including apparent 'monojets')~\cite{CasalderreySolana:2010eh}. This is a very substantial energy loss, but compatible with RHIC data after scaling the opacity with the expected temperature dependence $T^3$~\cite{Qin:2010mn}.

The azimuthal angle between the dijets is shown in CMS data in Fig.~\ref{figpb7}b~\cite{Chatrchyan:2011sx}, for both pp (top right) and different centrality ranges in Pb--Pb. Unlike the energy asymmetry, the shape of the angular correlation does not change appreciably with centrality and the two jets stay essentially back-to-back in azimuth (the increase in the baseline at small relative angles, visible in the most central bins, arises mostly from random background fluctuations when one jet has been completely absorbed). CMS has also measured the distribution of energy around the jets with most of the lost energy appearing in very  low momentum fragments far away from the jet direction~\cite{Chatrchyan:2011sx}. Both results point to a jet quenching mechanism where the energy is radiated away via multiple, soft gluons, which in turn may even re-interact in the medium and lead to a further degradation of the energy. This is contrary to most pre-LHC quenching models which assumed energy loss to proceed via a few, relatively hard gluons which would stay close to the original jet direction and would lead to a significant angular broadening between the dijets.

\section{Summary and conclusion}

The LHC has entered the field of ultra-relativistic heavy ion physics with a very impressive first performance. With a wealth of results within months after the first collisions, the LHC heavy ion program has benefited from many years of preparation, a very strong and complementary set of detectors, and, last but not least, a decade of experience and progress made at RHIC. In many aspects there is a smooth evolution from RHIC to LHC, qualitatively similar but quantitatively different; a sign both that the field is mature enough to be predictable and that the initial conditions (e.g. the energy density) are sufficiently different at both machines so that comparing the results and model predictions will reveal more about the properties of the QGP than looking at either facility alone. There are already a few unexpected or surprising results with the first low luminosity run, while we have only started to explore the 'terra incognita' where the energy reach of LHC makes it unique. 

\
\

\end{document}